\documentclass[aps,twocolumn,groupedaddress,prb]{revtex4}
\usepackage{psfig}
\newcommand {\LiHoF}{$\mathrm{LiHoF_4}$}
\newcommand {\LiYHoF}{$\mathrm{LiY_{96\%}Ho_{4\%}F_4}$}
\newcommand {\LiYF}{$\mathrm{LiYF_4}$}

\newcommand {\Hethree}{$\mathrm{{}^3He}$}

\newcommand {\Ho}{$\mathrm{Ho^{3+}}$}
\newcommand {\Y}{$\mathrm{Y^{3+}}$}

\begin{document}

% You should use BibTeX and apsrev.bst for references
\bibliographystyle{/usr/share/texmf/bibtex/bst/revtex4/apsrev}

%\preprint{}

\title{Single ion scattering contributions to
  the thermal conductivity of $\mathbf{LiHoF_4}$ and $\mathbf{LiY_{96\%}Ho_{4\%}F_4}$}
% Optional argument for running titles on pages
%\title[]{}


\author{James Nikkel}
\altaffiliation{Present address: Yale University, Sloane Physics Lab, New Haven, CT}
\author{Brett Ellman}
\email{bellman@kent.edu}
%\thanks{}
\affiliation{Kent State University, 105 Smith Laboratory, Kent, OH}


\date{\today}

\begin{abstract}
We have performed extensive zero-field thermal conductivity
measurements on single crystal samples of \LiHoF~ and \LiYHoF~
below $2.3~K$.  By comparing these data to a single ion
scattering model, we have shown that the thermal conductivity
of \LiYHoF~ is dominated by simple single-ion scattering while that of
\LiHoF~ shows additional contributions, possibly associated with collective
spin excitations.  No entirely satisfactory model, however, is available to
explain the thermal conductivity of the ferromagnet.
\end{abstract}

%\keywords{}

\maketitle

\section{Introduction}

\LiHoF~ has been the focus of
numerous experimental and theoretical studies ~\cite{reich:87, reich:90}.  The crystal field
constrains the angular momentum of the \Ho ions to lie strictly along the c-axis.  Furthermore,
the large Ho-Ho spacing greatly reduces the exchange term, making the system an excellent
approximation to a model dipolar Ising ferromagnet.  The non-magnetic ion \Y substitutes isostructurally
and isoelectronically
for \Ho, leading to dilute 3-d Ising systems.  The latter are, for low magnetic ion concentrations, known
to form a puzzling glassy system at low temperatures, whose signature is a {\it narrowing} distribution
of barriers to relaxation as temperature is lowered.  One goal of measuring thermal conductivity in these systems
is to obtain coarse phonon spectroscopic data to probe the 
\Ho~ ion energy levels in the low-temperature glass and ferromagnet.  Indeed, the \LiHoF~ and \LiYHoF~
characteristic hyperfine and magnetic energy level spectra
are measured in the 100's of milli-Kelvins, making thermal phonons at temperatures $\sim 1K$ an experimentally simple
way to access the characteristic energies of the system (simple, that is, when compared to, e.g., millimeter
photon spectroscopy).  The central questions we ask are (a) whether low-temperature collective excitations exist in the
Ising ferromagnet \LiHoF  below $T_c = 1.53 K$ and (b) whether the anomalous glassy behavior
of the dilute system \LiYHoF is due to rearrangements of correlated spin clusters.  In order to address these issues,
we must analyze the data in such a way as to deconvolve the rather complex
{\it single \Ho ion} contribution to the thermal resistance.  We will
accomplish this via a detailed analysis of the crystal field splittings and
resulting single-ion excitations that cause phonon scattering. 

%%%%%%%%%%%%%%%%%%%%%%%%%%%%%%%%%%%%%%%%%%%%%%%%%%%%%%%%%%%%%%%%%%%%%%%%%%%%% Theory 
\section{Scattering Model}

We start with the general form of the thermal
conductivity, $\kappa$, in terms of the phonon
heat capacity, $C({\bf k})$, the mean free
path, $l({\bf k})$, and the the speed of sound, $v({\bf k})$,
written as an integral over the
phonon wave vector, ${\bf k}$:

\begin{equation}
  \kappa = \int C({\bf k}) v({\bf k}) l({\bf k}) \, d^3 {\bf k}.
  \label{eq:kappa_prime_k}
\end{equation}
Our initial approximations are to assume that the system is structurally isotropic
with respect to ${\bf k}$ and that only acoustic phonons with a linear dispersion
relation
contribute at the experimental temperatures.
We can then calculate the thermal conductivity in terms of
an integral over the phonon frequency, $\omega$, as

\begin{equation}
  \kappa = \frac{4 \pi}{v^2} \int C(\omega) l(\omega) \omega^2 \, d \omega.
  \label{eq:kappa_prime_w}
\end{equation}

The specific heat, $C(\omega)$ contribution from phonons with
energy between $\omega$ and $\omega+d \omega$, is the
derivative of the mean thermal energy 
density of the lattice~\cite{klemens:58},

\begin{equation}
  C(\omega)\,d\omega = \frac{d\omega}{8 \pi^3} \frac{\partial}{\partial T} \frac{\hbar \omega}{e^{\hbar \omega / k_B T}-1}.
\end{equation}

Using equation~\ref{eq:kappa_prime_w}, the thermal conductivity is now

\begin{equation}
  \kappa = \frac{\hbar^2}{2 \pi^2 v^2 k_B T^2} \int_0^\infty \frac{\omega^4 e^{\hbar \omega \beta} l(\omega) \,d\omega}{(e^{\hbar \omega \beta}-1)^2}.
  \label{eq:kappa_int}
\end{equation}

Our task is to find $l(\omega)$, taking into account both single-ion magnetic
scattering and scattering due to, e.g., crystalline defects.
For reference, we reproduce in table~\ref{tab:kappa_l} a synopsis~\cite{klemens:58}
of the $\omega$ dependence of various structural scattering
mechanisms.

\begin{table}
  \centerline{
    \begin{tabular}[b]{|l|c|c|}
      \hline
      Type of scattering & $ l(\omega) \propto $  &  $ \kappa(T) \propto$ \\
      \hline
      External boundaries  &   $const. $   &  $T^3$      \\
      Grain  boundaries    &   $const. $   &  $T^3$      \\
      Stacking faults      &   $\omega^{-2}$  &  $T$      \\
      Conduction electrons &   $\omega$  &  $T^2$      \\
      Point defects        &   $\omega^{-4}$  &  $1/T$      \\
      Umklapp processes    &   $\omega$  &  $T^3 e^{\alpha/T}$      \\
      \hline
    \end{tabular}
  }
  \label{tab:kappa_l}
  \caption[Table of mean free path dependencies]
      {Table of the functional form of the mean free path, $l$, on $\omega$.} 
\end{table}
We also assume the correctness of Mathiesen's rule, i.e.,
that contributions the mean free path
are added reciprocally.

%%%%%%%%%%%%%%%%%%%%%%%%%%%%%%%%%%%%%%%%%%%%%%%%%%%%%%%%%%%%  Numerical integral Section
\section{Detailed single ion calculation}

The thermal resistance of an isolated \Ho ion may be
numerically estimated from the electronic transition
probabilities of the ions when they
are excited by thermal phonons.  We will not attempt here to calculate
from first principles the electron-phonon coupling constants.  Instead,
we will use a semi-phenomenological model to ascertain whether a single-ion
scattering term can account for the observed thermal conductivity.
The ${}^5I_8$ multiplet of 
the \Ho~ ions in \LiHoF~ is comprised of a ground state
doublet followed by two singlets at about 9 and $39~K$.
The energy levels above these are of sufficiently
high energy that we can neglect them at the experimental temperatures. 
The ground state doublets are each split into 8 hyperfine sublevels
which in turn may be split relative 
to each other due to a Zeeman interaction with an external or 
internal magnetic field.
The magnitude of the ground state magnetic splitting in the ferromagnet
was measured by Battison \textit{et al.}
~\cite{battison:75}.  Using 
an optical spectroscopic technique, they found it to be
proportional to the magnetization (i.e., mean field) with a maximum at 
zero temperature equal to $2.6~cm^{-1}$, or $3.7~K$.  

The energy levels are calculated by solving the eigenvalue
problem for the ${}^5I_8$ single ion Hamiltonian.  We follow the method described
by Giraud \textit{et al.} \cite{giraud:01}.
In the $|J,m,I,n\rangle$  basis, the Hamiltonian can be approximated as 
the sum of the crystal field, 
Zeeman, and hyperfine contributions,

\begin{equation}
  H = H_{CF} + H_{Zee} + H_{hf}.
\end{equation}
The crystal field contribution is written in term of the 
Stevens' operators, $O^m_l$, and the crystal field parameters, $B^m_l$.
The resulting crystal field Hamiltonian is 

\begin{eqnarray}
  H_{CF} &=& \alpha B^0_2 O^0_2 \\ 
  &+& \beta \left(B^0_4 O^0_4 + B^4_4 O^4_4 \right) \nonumber \\ \nonumber
  &+& \gamma \left(B^0_6 O^0_6 + B^4_6 O^4_6 \right), \nonumber
\end{eqnarray}
where 
$\alpha$, $\beta$, and $\gamma$ are ion specific 
coefficients calculated by Stevens~\cite{stevens:52}.
For \Ho, $\alpha = -1/450$, $\beta = -1/30030$, and $\gamma = -1/3864861$.
(The general form for $\alpha$ and $\beta$ can be found on page~253 of
Hutchings~\cite{hutchings:64}.)
The crystal field parameters, $B^m_l$, are given
by Gifeisman \textit{et al.} in 
reference~\cite{gifeisman:78} and were measured through
high resolution optical spectroscopy.  Their
values are: $B^0_2 = 273.9~K$, $B^0_4 = -97.7~K$,
$B^0_6 = -6.5~K$, $B^4_4 = -1289.1~K$, and
$B^4_6 = -631.6~K$.

\begin{table}
  \centerline{
    \begin{tabular}[b]{|c|l|}
      \hline
      $O^m_l$ & in $|J,m \rangle$  basis \\
      \hline
      $O^0_2$ & ~$3 J_z^2 - J(J+1)$ \\
      $O^0_4$ & ~$35J_z^4 - 30J(J+1)J_z^2 + 25J_z^2$ \\
      {}      & ~$~~ - 6J(J+1) + 3J^2(J+1)^2$ \\
      $O^4_4$ & ~$\frac{1}{2} \left[J_+^4+J_-^4 \right] $ \\
      $O^0_6$ & ~$231J_z^6 - 315 J(J+1)J_z^4$ \\
      {}      & ~$~~ + 735 J_z^4 + 105J^2(J+1)^2J_z^2$ \\
      {}      & ~$~~ - 525J(J+1)J_z^2 + 294J_z^2$ \\
      {}      & ~$~~ - 5J^3(J+1)^3 + 40J^2(J+1)^2$ \\
      {}      & ~$~~ - 60J(J+1)$ \\
      $O^4_6$ & ~$\frac{1}{4} \left[ \left( 11J_z^2 - J(J+1) - 38 \right) \left( J_+^4 + J_-^4 \right) \right. $ \\
      {}      & ~$~~ + \left. \left( J_+^4 + J_-^4 \right) \left( 11J_z^2 - J(J+1) - 38 \right) \right] $ \\
      \hline
    \end{tabular}
  }
  \label{tab:stev_op}
  \caption[Table of Stevens' operators]
      {Table of the Stevens' operators and equivalents.} 
\end{table}
The Stevens' operators, $O^m_l$ are given in reference~\onlinecite{stevens:52}
and~\onlinecite{hutchings:64} and the ones of interest 
here are summarized in table~\ref{tab:stev_op}.

The Zeeman contribution to the Hamiltonian is simply
\begin{equation}
  H_{Zee} = - g_J \mu_B \vec{J} \cdot \vec{H}
\end{equation}
where the g-factor $g_J=5/4$ for the $J=8$ ground
state, $\mu_B$ is the Bohr magneton, and $\vec{H}$ is
the magnetic field due to applied external fields and
the internal field in the ferromagnetic state.
Due to the symmetry of the crystal, the
internal magnetic field due to ferromagnetic order is directed along the
(Ising) $c-axis$. Therefore, in the absence of an external 
field, and within the mean-field approximation, we may write
\begin{equation}
  H_{Zee} = - g_J \mu_B \frac{H_0}{M_0} M(T) J_z,
  \label{eq:kappa_H_zee}
\end{equation}
where $M_0$ is the saturation magnetization at
zero temperature, $H_0$ is the effective maximum 
internal field at zero temperature, and
$M(T)$ is the temperature dependent magnetization.
Note that this is the {\it only} term in the Hamiltonian where 
the temperature enters into the calculation of the energy spectrum.
(The thermal broadening of energy levels described below also depends
on T.)
The magnetization of \LiHoF~ is calculated from a 
fit to the magnetization data measure in reference~\onlinecite{griffin:80}.
The zero temperature
internal field, $H_0$, is approximately $0.33~T$ which gives the
correct splitting of $3.7~K$ at zero temperature
as measured by Battison \textit{et al.}~\onlinecite{battison:75}.  Note that since our
sample geometry is a long bar along the c-axis, we do not apply a demagnetization correction.

The hyperfine contribution may be written
\begin{equation}
  H_{hf} = A_J \vec{J} \cdot \vec{I}.
  \label{eq:kappa_H_hf}
\end{equation}
The coupling parameter $A_J$ was found to be approximately $39~mK$ by Mennenga~\textit{et al.}~\cite{mennenga:84b}
for \LiHoF~ from specific heat measurements.  Magari\~{n}o~\textit{et al.}~\cite{magarino:76} 
found approximately the same value for 2\% \Ho~ diluted into \LiYF~ using electron paramagnetic resonance.

\begin{figure}[ht]
  \centerline{
    \hbox{\psfig{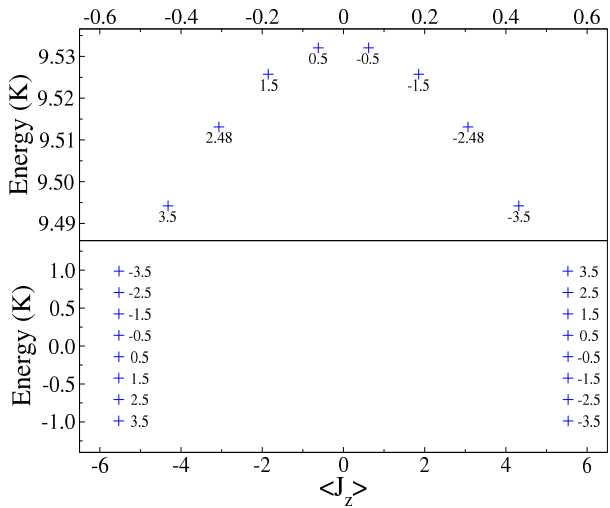}}}
  \caption[Energy eigenvalues, $\langle I_z \rangle$, and $\langle J_z \rangle$] %
	  {Plot of the energy eigenvalues arranged in terms of the $\langle J_z \rangle$ %
	    expectation values.  The labels on the points are the $\langle I_z \rangle$ %
	    expectation values.  The bottom plot shows the ground state doublet, and the top plot %
	    is the $1^{st}$ excited state.}
          \label{fig:eigen_highT}
\end{figure}

\begin{figure}[ht]
  \centerline{
    \hbox{\psfig{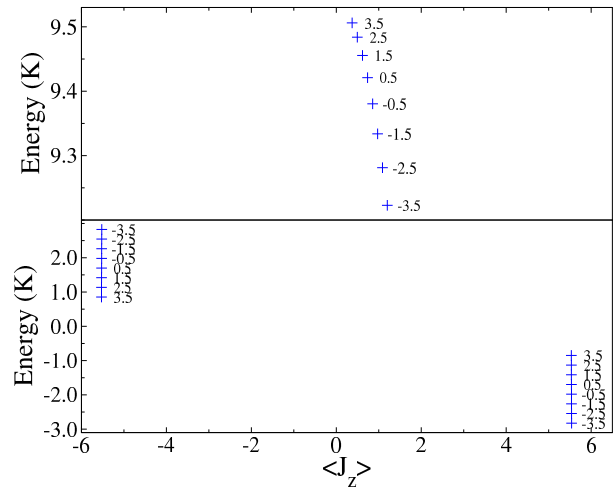}}}
  \caption[Energy eigenvalues, $\langle I_z \rangle$, and $\langle J_z \rangle$ with $M(T)$ splitting] %
	  {Similar to figure~\ref{fig:eigen_highT} but with magnetic splitting equal to %
	    what would occur in \LiHoF~ at $0~K$.  Note that the scale of the $\langle J_z \rangle$ axis is the %
	  same in the top and bottom of the plot.}
          \label{fig:eigen_lowT}
\end{figure}

For the \Ho~ ions in \LiHoF~ and \LiYHoF, $J=8$ and
$I=7/2$.  Therefore, the Hamiltonian takes the
form of a $136 \times 136$ Hermitian matrix.
Numerical diagonalization gives the respective eigenvectors, which
in turn allow the calculation of pertinent expectation values.
Specifically, it is important to know the expectation value of
the $z$-component of the nuclear angular momentum,
$\langle I_z \rangle$,
because transitions that change $\langle I_z \rangle$
a large amount are strongly suppressed.
Figure~\ref{fig:eigen_highT} is a plot of the ground state doublet and the
first excited state eigenvalues plotted against the expectation values
$\langle J_z \rangle$.  The value of $\langle I_z \rangle$ is written next to 
each point.
The values for $\langle I_z \rangle$ are uniformly half integer as 
we would expect.  Note that the scale of the $\langle J_z \rangle$ axis
on the top of the plot is one tenth that of the bottom to show the form of
the excited state more clearly.
Figure~\ref{fig:eigen_lowT} is similar to figure~\ref{fig:eigen_highT}
but with a magnetic field turned on in the $z$ direction.  
The field used here is about $0.3~T$, equal to that which would
be generate internally by \LiHoF~ at zero temperature.  As we will discuss,
only transitions between energy levels with $\Delta I = 0$ will be included
in our computations.

There is one important note here about the energy eigenvalues
in our particular case.  The crystal field parameters measured by
Gifeisman \textit{et al.}~\cite{gifeisman:78} were measured through
high resolution optical spectroscopy and they did \textit{not}
observe the $9.5~K$ excited state in their luminescence spectra.  
Their parameters inserted into the calculation therfore predict 
that this unconstrained level sits at $14.5~K$ above the ground state doublet.  
Since this is the only state so effected, we adjust our eigenvalues 
by subtracting $5~K$ from this state before using them in further 
calculations.

Recall that the goal is to
calculate a temperature dependent mean free path, $l(\omega)$, 
that is to be inserted in to the 
integral in equation~\ref{eq:kappa_int}.
In the present case,
we assume that there are only three
scattering processes, giving us three independent
$l$'s to add. 
The first is due to phonon
scattering from the sample boundaries. This term is 
simply the geometrically averaged sample size, $l_{br}$, and is independent of $\omega$.
The second term is from scattering off of point defects, $l_{pd}$, which
gives a term proportional to $1/ \omega^4$ (see table~\ref{tab:kappa_l}).
As we shall see, this term is
necessary to obtain agreement with the high 
temperature range of our data (though higher dimensional defects might also suffice, we
assume point defects for simplicity).
The last process is due to the non-trivial scattering from
the \Ho~ ions, $l_{ion}$.  Therefore,

\begin{equation}
  l=\frac{1}{\frac{1}{l_{br}}+\frac{1}{l_{pd}} +\frac{1}{l_{ion}}}.
\end{equation}
We choose to write the mean free path due to the single ion scattering
process as $l_{ion}=l_{min}/P_{scatt}$.  The scale factor $l_{min}$
is similar to the minimum distance that a phonon can go without
running into a \Ho~ ion, and $P_{scatt}$ is the
probability that the phonon will scatter off of that
ion.  We define $l_{pd} \equiv \frac{\xi}{\omega^4}$, 
and will use $\xi$ as a fitting parameter.  
Our complete expression for the mean free path is then

\begin{equation}
  l(\omega)=\frac{1}{\frac{1}{l_{br}} + \frac{\omega^4}{\xi} + \frac{P_{scatt}(\omega,T)}{l_{min}}}.
  \label{eq:kappa_total_mfp}
\end{equation}

After computing the energy level spectrum at a particular temperature,
we calculate the partition function which
is obtained by simply summing over each level, $E_i$,
integrated over the level density, $D_i(\epsilon)$,

\begin{equation}
  Z= \sum_{i} \int_{-\infty}^{\infty} e^{-\beta \epsilon}D_i(\epsilon)\,d\epsilon.
\end{equation}
We assume that the individual energy levels
are thermally broadened (due to crystal field fluctuations from
thermal motion of the atomic neighbors of the \Ho ions in the lattice) with a Gaussian energy distribution.
For simplicity, we further assume that all of the levels are have the same
width, $\Delta E$, resulting in the following energy density 
function, 

\begin{equation}
  D_i(\epsilon) = \frac{1}{\Delta E \sqrt{\pi}\,}e^{-\frac{(\epsilon-E_i)^2}{\Delta E^2}}.
\end{equation}
Our partition function now becomes

\begin{equation}
  Z= e^{\frac{\beta^2 \Delta E^2}{4}} \sum_{i} e^{-\beta E_i}.
\end{equation}

The scattering probability, $P_{scatt}(E_P,T)$, is 
a function of the phonon energy, $E_P=\hbar \omega$,
and the temperature of the system, $T$.
We calculate it by performing  a double sum over all the energy levels,
calculating the scattering probabilities between each pair.

\begin{eqnarray}
  \lefteqn{P_{scatt}(E_P,T) = \delta \left( \left| \langle I_z^{(i)} \rangle - \langle I_z^{(j)} \rangle \right| \right) } \\
   && \times \sum_{i,j} \left[ \int_{-\infty}^{\infty}e^{-\beta \epsilon}D_{ij}(\epsilon)\,d\epsilon \right. \nonumber\\
    && ~~- \left. \int_{-\infty}^{\infty}e^{-\beta \epsilon}D_{ji}(\epsilon)\,d\epsilon \right] \nonumber,
  \label{eq:p_scatt_eq}
\end{eqnarray}
where $D_{ij}(\epsilon)=D_{i}(\epsilon) \, D_{j}(\epsilon+E_P)$ and
$D_{ji}(\epsilon)=D_{j}(\epsilon) \, D_{i}(\epsilon-E_p)$.  This particular
form takes into account stimulated emission from the higher level to the
lower one as well as the direct excitation of the ion.  The delta function 
allows for excitations only between levels that have the same
nuclear angular momentum expectation values (see Fig's \ref{fig:eigen_highT} and
\ref{fig:eigen_lowT}).  Slightly decreasing this selectivity, 
say by allowing $\left| \langle I_z^{(i)} \rangle - \langle I_z^{(j)} \rangle \right|$
to vary by one or two, makes little difference to the current model
except at the lowest temperatures ($<100~mK$). 
Using the level density mentioned above, the scattering probability 
is found to be

\begin{eqnarray}
  \lefteqn{P_{scatt}(E_P,T) =  \frac{1}{Z} e^{\frac{1}{8}\beta^2 \Delta E^2}} \\
  && \times \left\{ e^{\beta E_P/2} - e^{-\beta E_P/2} \right\} \nonumber\\
  && \times \sum_{i,j} e^{-\frac{(E_i-E_j+E_P)^2}{2 \Delta E^2}} e^{-\frac{1}{2}\beta^2 (E_i+E_j)} \nonumber \\
  && \times \delta \left( \left| \langle I_z^{(i)} \rangle - \langle I_z^{(j)} \rangle \right| \right) \nonumber.
  \label{eq:p_scatt_eq_final}
\end{eqnarray}

The energy level width,  $\Delta E$, requires attention here.
The thermal energy due to deformation of the lattice is essentially
 
\begin{equation}
  E_T = k_b T = \frac{1}{2} C_0 \epsilon^2,
\end{equation}
where $C_0$ is an averaged elastic constant, and $\epsilon$
is an averaged strain.
For small lattice deformations
changes in the crystal field energy are therefore proportional to the strain, 
\begin{equation}
  \delta E_{CF} \propto  \epsilon.
\end{equation}
We collect the constants, and designate the thermal
broadening as 

\begin{equation}
 \Delta E = A \sqrt{T}.
 \label{eq:kappa_delta_E}
\end{equation}
with $A$ as a fitting parameter. 

\begin{figure}[ht]
  \centerline{
    \hbox{\psfig{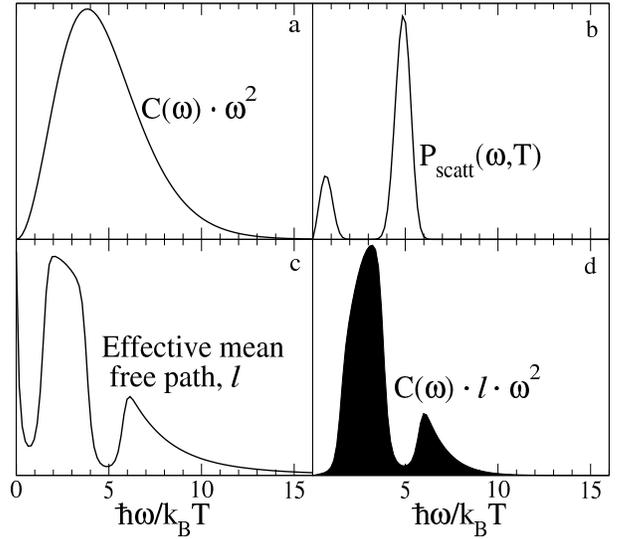}}}
  \caption[Integrand elements for $\kappa$] %
	  {Plots of the contributions to the heat capacity calculation for $T=2~K$.  The various
	  panels are described in the text.}
          \label{fig:kappa_scatt_terms}
\end{figure}

The thermal conductivity is calculated at each temperature by the following
procedure.  First the various temperature dependent parameters (internal field for \LiHoF and
the broadening parameter A) are 
set and the Hamiltonian eigenvalue problem is solved.  The integral in equation~\ref{eq:kappa_int}
is computed by inserting the mean free path, $l(\omega)$,  from equation~\ref{eq:kappa_total_mfp}.
The scattering probability is then calculated at each point in the integral
using equation~\ref{eq:p_scatt_eq} where $E_P=\hbar \omega$.

We have used an integration cut-off of $\omega = 16 k_B T /\hbar$, which is quite sufficient as
seen in Figure~\ref{fig:kappa_scatt_terms}. 
Panel~$a$ is the heat capacity part of the integrand which
essentially vanishes for higher frequencies.
Panel~$b$ is the scattering probability from equation~\ref{eq:p_scatt_eq}, 
and panel~$c$ is $l(\omega)$ from equation~\ref{eq:kappa_total_mfp}.
The last panel, $d$, is the integrand to equation~\ref{eq:kappa_int}, 
so the area of the filled portion represents the total thermal
conductivity at this temperature.  The effect of transitions between \Ho levels
is quite apparent in the mean free path data, effectively removing specific
frequency bands from the thermal conductivity integral.

%%%%%%%%%%%%%%%%%%%%%%%%%%%%%%%%%%%%%%%%%%%%%%%%%%%%   Experimental stuff
\section{Experimental Details}
\label{sec:kappa_exp}

Thermal conductivity was measured by the standard gradient method using a heater
and two thermometers.
Our heaters were $200~\Omega$ metal film on Kapton strain gauges and the
thermometers used were $1~k\Omega$ to $10~k\Omega$ 
$\mathrm{RuO_2}$ surface mount resistors, depending on the temperature range of interest.
Copper bars approximately $150~\mu m$ thick were silver epoxied to
the sample dividing it approximately into thirds.  These bars defined the
distance over which the temperature gradient, $\Delta T=T_2-T_1$,
was measured.  The heaters were wired with superconducting
NbTi wires to eliminate ohmic heating and to minimize heat leaks, while
thin Evanohm wires were used on the thermometers.
To check for thermal leakage through the wires, the measurement
of \LiYHoF~ was repeated with the wires doubled in length.
The results of the thermal conductivity between the two sets of measurements
were within the noise of either experiment.
Our two samples (pure and dilute) were optically oriented and cut such that
the long length was along the $c-axis$.  The pure \LiHoF~
was $1.46 \times 1.56 \times 5.4~mm^3$.  This gives a
cross section of $0.0228~cm^2$.  The distance, $L$, between the 
thermometers was $0.22~cm$.
The $11.7~mg$ dilute \LiYHoF~ sample was $0.93 \times 0.75 \times 4.7~mm^3$,
but slightly trapezoidal.  The effective cross section was calculated
by using measurements from a digital photograph of the sample.  The
final value used was $A=0.0066~cm^2$.
The distance between thermometers for this sample was also $0.22~cm$.
Most experiments were conducted in a $^3He$ cryostat.  Data below
$310 mK$ were obtained in a helium dilution refrigerator.

%%%%%%%%%%%%%%%%%%%%%%%%%%%%%%%%%%%%%%%%%%%%%5 Results
\section{Experimental Results}

%%%%%%%%%%%%%%%%%%%%%%%%%%%%%%%%%%%%% Pure
\subsection{Pure \LiHoF}
\label{sec:kappa_exp_pure}

\begin{figure}[ht]
  \centerline{
    \hbox{\psfig{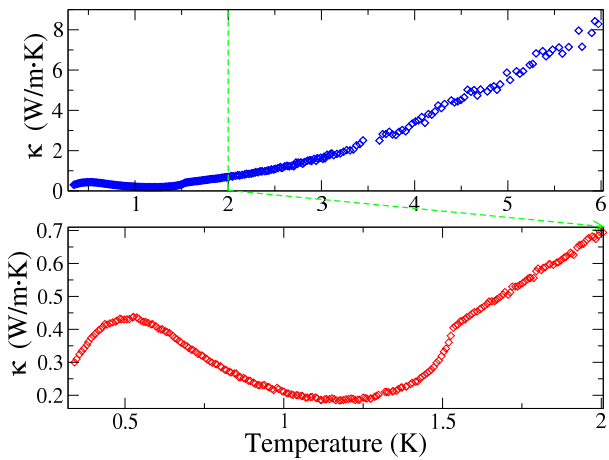}}}
          \caption[$\kappa$ vs. $T$ for \LiHoF]
		  {Measured thermal conductivity, $\kappa$,  for \LiHoF.  The bottom
		  panel is a magnification of the top to better illustrate the
		  structure below the Curie Temperature, $T_C$.}
          \label{fig:kappa_pure_res}
\end{figure}

The accumulated data for a number of experiments are plotted
in figure~\ref{fig:kappa_pure_res}.  The top
panel is the entire range of data obtained, and
the bottom is a magnification of data below $2~K$.
One can easily see the ferromagnetic transition on this plot at 
$\approx 1.54~K$.  One interpretation of the overall structure of the data is as
follows;  as we go down in temperature, the thermal conductivity
decreases due to the $T^3$ term in the specific heat.
At this point, the vast majority of phonon-driven 
transitions are between the hyperfine ground state sublevels.
When we reach $T_C$, the ground states start Zeeman splitting, and there are
a number of new transition possibilities opened up, so the conductivity drops
rapidly.  Note that the peak in $C(\omega)$ in figure~\ref{fig:kappa_scatt_terms}
occurs at approximately $\omega=4 k_b T /\hbar$.  This implies that at $T_C$, about
$1.5~K$, the majority of phonons have energies around $6~K$.  However, 
just above $T_C$, the \Ho energy spectrum looks like that shown in 
figure~\ref{fig:eigen_highT}, where there are no scattering possibilities for phonons
between about $2~K$ and about $8.5~K$.  Below $T_C$, the
excluded temperature region quickly shrinks due to the splitting, and
simultaneously, the peak of the phonon spectrum decreases.  
At about $1.2~K$, the ground states have split enough that 
transitions are suppressed and the conductivity starts increasing.  
Eventually, this rise is expected to be countered by the $T^3$ factor from having a sample
of finite size.
While this qualitative explanation was the primary motivation to develop a single ion
model of the thermal conductivity we shall see that this picture is not 
sufficient to explain all of the data.
\begin{figure}[ht]
  \centerline{
    \hbox{\psfig{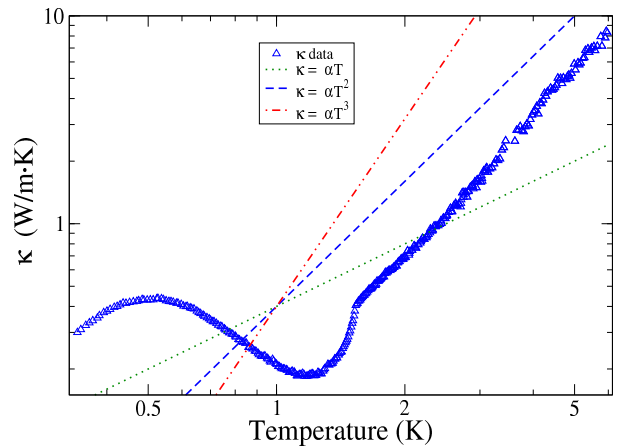}}}
  \caption[log-log $\kappa$ vs. $T$ for \LiHoF]
	  {log-log plot of $\kappa$ for \LiHoF~ along with some lines of $T^n$ for comparison
		    with table~\ref{tab:kappa_l}.}
          \label{fig:kappa_pure_res_log}
\end{figure}

Figure~\ref{fig:kappa_pure_res_log} shows the whole range of data on a log-log plot,
along with lines proportional to $T$, $T^2$, and $T^3$.
Recall from table~\ref{tab:kappa_l} the various structural scattering processes that can
contribute to the thermal conductivity.
As one can see, the data do not follow any one of these power
laws over any significant temperature range.  We did not take measurements
of the pure \LiHoF in the dilution refrigerator, so we can not see
the approach to $T^3$ in the low temperature limit.

%%%%%%%%%%%%%%%%%%%%%%%%%%%%%%%%%%%%%%%%%%%%%% Dilute
\subsection{Dilute \LiYHoF}

\begin{figure}[t]
  \centerline{
    \hbox{\psfig{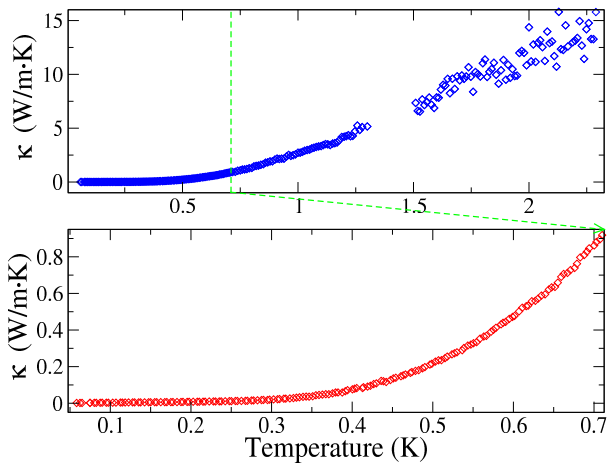}}}
          \caption[$\kappa$ vs. $T$ for \LiYHoF]
	      {Measured $\kappa$ for \LiYHoF.  The bottom panel is 
		a magnified view of the top.}
          \label{fig:kappa_dilute_res}
\end{figure}

We have again plotted our data in two temperature ranges
in figure~\ref{fig:kappa_dilute_res}.  This plot is
not very illuminating by itself since it
goes so quickly and smoothly to zero.  
There is no ferromagnetic transition, so there are no sharp 
structures as there are for \LiHoF.  (The gap in the
data between $1.3~K$ and $1.5~K$ is due to T-control issues
in the \Hethree~ cryostat.)
As in the previous section, the data becomes increasingly noisy at higher
temperatures due to the reduced sensitivity of
the thermometers and the rapidly increasing thermal
conductivity. Note that at $2~K$ the thermal
conductivity of \LiYHoF~ is about 10 times that 
of \LiHoF, but slightly smaller at $0.5~K$.

\begin{figure}[ht]
  \centerline{
    \hbox{\psfig{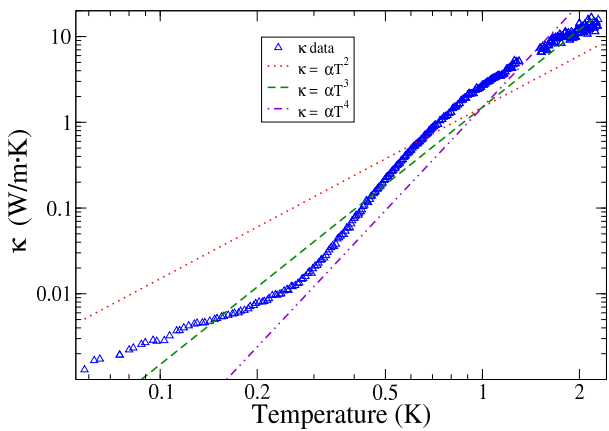}}}
          \caption[log-log $\kappa$ vs. $T$ for \LiYHoF]
	      {log-log plot of $\kappa$ for \LiYHoF~ along with some $T^n$ 
		lines for comparison.}
          \label{fig:kappa_dilute_res_log}
\end{figure}

Figure~\ref{fig:kappa_dilute_res_log} is a plot of 
$\log(\kappa)$ versus $\log(T)$ to check for regions
where the thermal conductivity is dominated by a single
power law.  It appears that at the lowest temperatures the
data may be approaching $T^3$, but going to lower
temperatures would be necessary to confirm this.
More importantly, the log plot brings out
structures that must be reproduced by our single-ion model
if it is to describe the data.

%%%%%%%%%%%%%%%%%%%%%%%%%%%%%%%%%%%%%%%%%%%%   Analysis
\section{Analysis}
\label{sec:kappa_analysis}

Our goal is to find a set of parameters that provide the
best fit for both our \LiHoF and \LiYHoF~ data \textit{simultaneously}.
Fitting parameters include
$\xi$, which determines the contribution from point
defect scattering, and $l_{min}$, which determines the size of
the contribution from the \Ho~ ion scattering.
$P_{scatt}$ also contains a
parameter that can be varied, as does the
the thermal broadening, Eq. ~\ref{eq:kappa_delta_E}.
Since $A$ is a single ion property, it
should be independent of the concentration of the \Ho~
ions.  Similarly, $l_{min}$ should be proportional to the
inverse of the cube root of the
concentration, fixing its relative value.
The effective mean free path due to the finite
size of the sample, $l_{br}$, is calculated from
the sample dimensions and is fixed.
We take $A_J$ from equation~\ref{eq:kappa_H_hf}
to be a constant, $38.6~mK$, since this value
has been measured and published using at least two
different methods~\cite{mennenga:84b, magarino:76}.  
As mentioned, $H_0 \approx 0.33 T$ from equation~\ref{eq:kappa_H_zee} 
was indirectly, and somewhat imprecisely, measured by Battison \textit{et al.}~\cite{battison:75}.

We will first discuss the dilute glassy system \LiYHoF~.

%%%%%%%%%%%%%%%%%%%%%%%%%%%%%%%%%%%%%%%%% Dilute
\subsection{Dilute \LiYHoF}

\label{sec:kappa_analysis_dilute}
\begin{figure}[ht]
  \centerline{
    \hbox{\psfig{figure=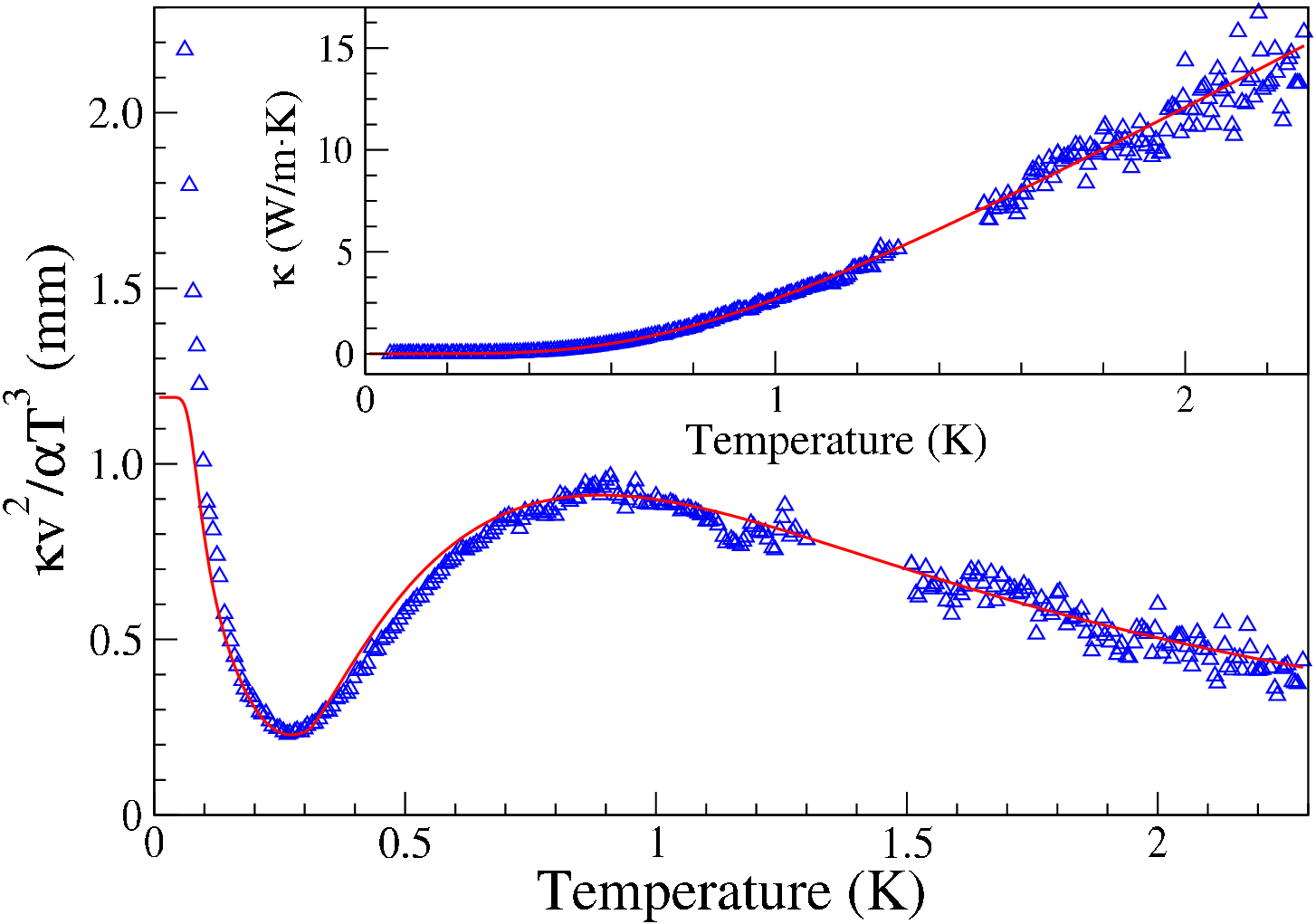,width=8cm,%
                clip=}}}
          \caption[Best single ion model $\kappa$ fit of \LiYHoF]
		  {Plot of $\kappa v^2 / \alpha T^3$ and best fit for \LiYHoF.  Inset is 
		  $\kappa$.}
          \label{fig:kappa_dilute_fit_T3}
\end{figure}

Figure~\ref{fig:kappa_dilute_fit_T3} is a plot
of $\kappa$ and $\kappa v^2 / \alpha T^3$
for \LiYHoF~ along with the best model fit. Here, 

\begin{equation}
  \alpha = \frac{2 \pi^2}{15} \frac{k_b^4}{\hbar^3}
\end{equation}
which gives us $\kappa/T^3$ roughly in units of the mean free path (the
equivalence would be exact of $l$ did not depend on $\omega$).
As we can see, the fit is quite good with this parameter set.
The best fit parameters are
$A=0.45 \pm 0.03~K^{1/2}$, $l_{min}=38 \pm 1~\mu m$, and $\xi=15 \pm 2 ~m \cdot K^4$.
The uncertainties were estimated 
by varying the individual parameters to obtain the same minimum $\chi^2$,
and do not reflect any correlations between the parameters.
As we can see, the scattering due to point defects is a fairly
small contribution.  For example, a $1~K$ phonon will have a mean
free path, due to this process, of about 15 meters.  Only
for phonons with energies greater than about $12~K$
does the mean free path to become smaller than
the physical size of the sample.  Because of this, the initial fitting
was done with $\xi=\infty$, after which $\xi$ was adjusted to minimize
the $\chi^2$.

\begin{figure}[ht]
  \centerline{
    \hbox{\psfig{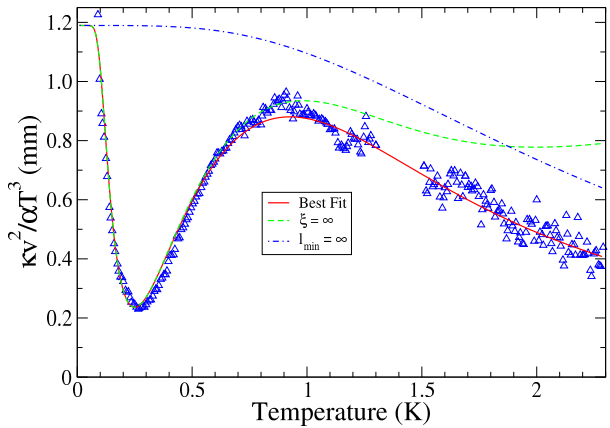}}}
          \caption[$\kappa v^2 / \alpha T^3$ and more fit variations for \LiYHoF]
		  {Plot of $\kappa v^2 / \alpha T^3$ and variations on the best fit for \LiYHoF. 
		  The parameters are described in the text.}
          \label{fig:kappa_dilute_fit_T3_anno2}
\end{figure}

Figure~\ref{fig:kappa_dilute_fit_T3_anno2} demonstrates how each
parameter effects the model.  Unless noted, all lines shown here use 
the ``best fit'' parameters.
The dashed line has the point defect scattering 
turned off by setting $\xi=\infty$.  Note that this line
follows the best fit line fairly well for temperatures
below the peak at $900~mK$, and is identical for temperatures
less than $500~mK$. 
The dot-dashed line demonstrates the effect of turning
off the \Ho~ single ion scattering process by by setting $l_{min}=\infty$.  
If we turned off the 
point defect scattering at the same time, we would obtain a 
horizontal line at $\kappa v^2 / \alpha T^3$=$l_{br}$=$1.19~mm$.
From this, we see that the single ion process alone is
responsible for the structure of the dip at $270~mK$ and
the addition of the point defect scattering causes a
turn-over in the data to produce the peak at $900~mK$.

In the very low temperature region, less than $80~mK$,
the model demonstrates a flattening behavior that
is not present in the data.  This is due to the value of
$l_{br}$, since it sets the overall maximum mean free path.
Interestingly, increasing $l_{br}$ beyond its physical
value does not improve the overall model, as this 
effectively scales the entire fit.
There currently is no parameter
set in with this model that reproduces the $T<80~mK$
region while retaining agreement with
the rest of the data set.  One explanation for this 
could be related to the surface condition of the sample.
At very low temperatures, phonons will have an increasing
wavelength, which will eventually become larger than the surface 
roughness.  When this occurs, the boundary scattering becomes
specular and the effective mean free path due to the finite size of
the sample increases.  This would result in a rapid upturn of
$\kappa /T^3$ when this limit is reached, consistent with our
measurements.

We believe that this single ion model
describes the thermal
conductivity of \LiYHoF~ fairly well below $2.5~K$,
and demonstrates the relative importance of
scattering from point defects.
We see no sign of the 
``cluster glass'' proposed by Reich \textit{et. al}~\cite{reich:90}.
They observed narrowing in the magnetic susceptibility 
with decreasing temperatures between $300~mK$ and $150~mK$,
but we see no signs of collective behavior in this temperature range.

%%%%%%%%%%%%%%%%%%%%%%%%%%%%%%%%%%%%%%%%%%% Pure
\subsection{Pure \LiHoF}
\label{sec:kappa_analysis_pure}

\begin{figure}[ht]
  \centerline{
    \hbox{\psfig{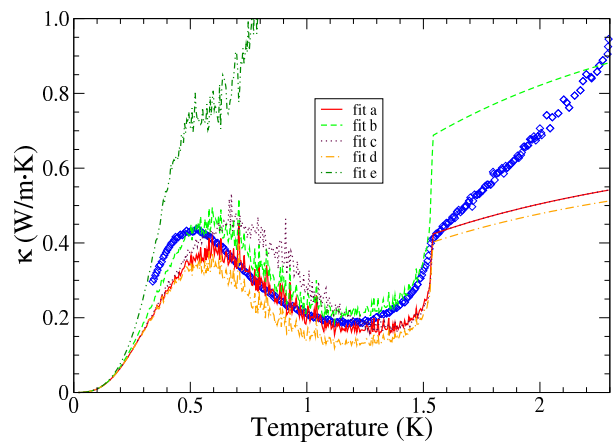}}}
  \caption[(Color online) Single ion model $\kappa v^2 / \alpha T^3$ fits of \LiHoF]
	  {Plot of $\kappa v^2 / \alpha T^3$ and three fits for \LiHoF.  
	    Parameters are described in the text.}
          \label{fig:kappa_pure_fit}
\end{figure}

We now move to the ferromagnet \LiHoF.
Because $A$ should be independent of
the concentration of the \Ho~ ions, and
$l_{min}$ should be proportional to the inverse of the cube root
of the concentration, 
we will start with $A=0.45~K^{1/2}$,
$l_{min}={}^3\sqrt{4\%} \cdot 38=13~\mu m$, and $\xi=15 ~m \cdot K^4$.
from the analysis of \LiYHoF.  We also turn on the Zeeman splitting 
by setting $H_0$ from equation~\ref{eq:kappa_H_zee} to $0.33~T$. 
Figure~\ref{fig:kappa_pure_fit}
is a plot of $\kappa v^2 / \alpha T^3$
versus temperature for \LiHoF~ with three lines from
our model.  Note that the vertical axis is on a log scale so that the structure
of the ferromagnetic transition can be seen.
The solid line is our ``best fit'' with the above parameters.
The dashed line uses the same parameters except for $\xi=\infty$
to see the effect of turning off the point defect scattering.
The dot-dashed line has $l_{min}=\infty$ to demonstrate the effect of
turning off the \Ho~ single ion scattering contribution (the individual
contributions are more profitably viewed using the color figure available
online).
As we can clearly see, the ``best fit'' is poor.
Furthermore, no amount of adjustment of the parameters
appears to yield significantly better agreement. 
We do, however, see some qualitative successes with this model
of \LiHoF.  The behavior immediately below the Curie
temperature is generally what we predicted 
from section~\ref{sec:kappa_exp_pure} and we
see a reasonably accurate roll off in $\kappa/T^3$ at the lowest temperatures.
One possibility
is that there are one or more additional scattering
mechanisms needed in the model.  For example, 
below the Curie temperature, domain formation 
could introduce a temperature dependent 
grain boundary-like scattering process.  However, this
would have no effect above $T_C$, where the deviation from the fit is most
pronounced.  (Of course, it is possible that difficulties present below $T_C$
in the model result in best-fit deviations {\it above} the Curie point.)
The second, and potentially most interesting possibility is that phonon-excited
spatially correlated spin-flips account for the enhanced scattering.  
Note that, unlike isotropic or nearly
isotropic systems, conventional spin-waves are not expected in an
ensemble of Ising spins.  Indeed, the nature of (potentially collective)
magnetic excitations above and below $T_C$ in this dipolar-coupled system are
unclear, as is the
question of whether such magnetic scattering could account for the behavior
we observe.
Again,
one would naively believe such effects to strongly dominate below $T_C$.
At higher temperature it is unclear what multi-spin excitations
are available to couple to the phonon current.

In conclusion, a detailed semi-phenomenological model of single-ion
phonon scattering processes is in excellent agreement with thermal conductivity
data on dilute \LiYHoF, with no evident signature of low-energy collective excitations involving random spin clusters.  Similar data on the pure ferromagnetic
system \LiHoF, however, are not in quantitative agreement with the independent
ion theory and will presumably require a study of collective effects in this
dipolar Ising system.
\bibliography{kappaj}
\end{document}